\documentclass[final,5p,times,twocolumn]{elsarticle}

\usepackage{amssymb}
\journal{Physics Letters B}

\begin{document}

\begin{frontmatter}



\title{Information loss and entropy conservation in quantum corrected Hawking radiation }


\author{Yi-Xin Chen}
\ead{yxchen@zimp.zju.edu.cn}
\author{Kai-Nan Shao}
\ead{shaokn@gmail.com}

\address{Zhejiang Institute of Modern
Physics, Zhejiang University, Hangzhou,
Zhejiang 310027, P.R.China}

\begin{abstract}

It was found in [Phys.Lett.B 675 (2009) 98] that information is conserved in the process of black hole evaporation,
 by using the tunneling formulism and considering the correlations between emitted particles. 
 In this Letter, we shall include quantum gravity effects, by taking into account of the log-area correction to Bekenstein-Hawking entropy.
The correlation between successively emitted particles is calculated, with Planck-scale corrections. 
By considering the black hole evaporation process, 
entropy conservation is checked, and the existence of black hole remnant is emphasized.
We conclude in this case information can leak out through the radiation and black hole evaporation is still a unitary process.

\end{abstract}


\end{frontmatter}


\def \w {\omega}
\def \k {\kappa}

\section{Introduction}

In 1975 Hawking discovered the remarkable fact that black holes radiate a thermal spectrum of particles 
and the temperature of this radiation depends on the surface gravity $\k$ of the black hole \cite{Hawking:1974sw}.
The discovery of temperature also gives connections between black holes and thermodynamics \cite{thermodynamics}. 
During black hole evaporation, all information about the original quantum state that formed the black hole seems to be lost, 
and a pure quantum state can evolve into a mixed one, thus violating the unitarity in quantum mechanics \cite{Hawking:1976ra}. 
Many attempts have been made to resolve the so-call information loss paradox \cite{Russo:2005aw}. 
Recently, using the non-thermal radiation spectrum obtained by tunneling formulism \cite{Srinivasan:1998ty,Parikh:1999mf,Shankaranarayanan:2000qv}, 
it is pointed out in \cite{Zhang:2009jn} that correlations exist among emitted particles, 
and information is leaked out through the radiation. The total entropy is conserved and the black hole evaporation process is unitary.
However, quantum gravity effects is not considered in resolving this paradox.
The black hole spectrum seen from an observer at infinity is dominated by modes that propagate from "near" the horizon 
where they have arbitrarily high frequencies and their wavelengths can easily go below Planck length \cite{Arzano:2005fu}. 
It is plausible that the motion of particle tunneling through the horizon might be affected by Planck-scale corrections. 
As a result, statistical correlation between quanta emitted and the fate of black hole in its late stages of evaporation are also influenced by quantum gravity effects. 
So it is essential to include quantum gravity corrections.
A modification of radiation spectrum, which includes Planck-scale corrections, is proposed in \cite{Arzano:2005rs}. 
In this letter we shall calculate the correlations in the situation of quantum gravity corrections.
We shall also check the conservation of entropy in radiation process.
We emphasize that the existence of black hole remnant is essential to preserve entropy conservation, after inclusion of quantum gravity corrections, 
while in classical cases black hole would evaporate completely.

Tunneling formulism has been proved as a relatively simple and straightforward method to calculate Hawking temperature and radiation spectrum \cite{Parikh:1999mf,temperatures}. 
In Section \ref{sec2} we review the key ingredients of this method and its close relation with black hole thermodynamics. 
The modified radiation probability is then obtained from the quantum gravity corrected black hole entropy.
 Using this probability, correlation between two successively emitted particles is calculated in Section \ref{sec3}.
 In section \ref{sec4} black hole evaporation is treated as a process of successive particles emitted out, 
and entropy carried by radiation is calculated. Furthermore, entropy conservation is checked and the existence of black hole remnant is discussed.

\section{\label{sec2} Tunneling, thermodynamics and modified radiation spectrum}

In this section we briefly review the tunneling formulism, 
and show that it is closely related to black hole thermodynamics \cite{Arzano:2005rs,Zhang:2008rd,Pilling:2007cn} .

Consider a general class of static, spherically symmetric spacetime
\begin{eqnarray}
\label{generalmetrics}
ds^2 = - A(r) dt_s \; ^2 + \frac{dr^2}{B(r)} + r^2 d\Omega^2,
\end{eqnarray}
$t_s$ is Schwarzschild time. The horizon is $r_H$, with $A(r_H) = B(r_H) = 0$. 
For Schwarzschild black hole, $A(r) = B(r) = 1 - \frac{2 M}{r}$ and $r_H = 2M$.
 This metric has a coordinate singularity at $r = r_H$, which can be removed by transforming to Painlev{\' e} coordinates
\begin{eqnarray*}
ds^2 = - A(r) dt^2 + 2 A(r) \sqrt{\frac{1-B(r)}{A(r) B(r)}} dt dr + dr^2 + r^2 d\Omega^2.
\end{eqnarray*}
with transformation $dt_s = dt - \sqrt{\frac{1 - B(r)}{A(r) B(r)}} dr$. 
The test particle is a massless spherical shell, which travels along radially null geodesics in this background
\begin{eqnarray}
\label{nullgeodesics0}
\frac{dr}{dt} \equiv \dot{r} = \sqrt{\frac{A(r)}{B(r)}}\left( \pm 1 - \sqrt{1-B(r)} \right) ,
\end{eqnarray}
where the positive (negative) sign gives outgoing  (incoming) radial geodesics.
Since $A(r)$, $B(r)$ are zero on the horizon, we can expand them as the following forms
\begin{eqnarray*}
A(r) &= A'(r_H) (r - r_H) + \mathcal{O}(r - r_H), \nonumber \\
B(r) &= B'(r_H) (r - r_H) + \mathcal{O}(r - r_H) .
\end{eqnarray*}
The surface gravity on the horizon is a Christoffel component for our choice of metric
\begin{eqnarray}
\label{surfacegravity}
\k = \Gamma^0_{ 0 0} = \frac{1}{2} \sqrt{\frac{1-B(r)}{A(r) B(r)}} B(r) \frac{d A(r)}{d r} |_{r = r_H}.
\end{eqnarray}
Near horizon, surface gravity can be expressed as
\begin{eqnarray}
\label{NHRsurfacegravity}
\k = \frac{1}{2} \sqrt{A'(r_H) B'(r_H)} + \mathcal{O}((r - r_H)^2).
\end{eqnarray}
and the null radial geodesics equation (\ref{nullgeodesics0}) is rewritten as
\begin{eqnarray}
\label{nullgeodesic}
\dot{r} & = & \frac{1}{2} \sqrt{A'(r_H) B'(r_H)} (r - r_H) + \mathcal{O}((r - r_H)^2) \nonumber \\
	& = & (r - r_H) \; \k.
\end{eqnarray}
The tunneling rate for particles through the event horizon is related to the imaginary part of the particle's action, $\Gamma \sim exp(-2 \textrm{Im} \mathcal{I})$, and
\begin{eqnarray}
\label{action1}
\textrm{Im} (\mathcal{I}) & = & \textrm{Im}  \int_{r_{in}}^{r_{out}} p_r dr = \textrm{Im} \int_{r_{in}}^{r_{out}} \int_{M}^{M-\w} \frac{d H'}{\dot{r}} dr \nonumber \\
	  & = & - \textrm{Im} \int_{r_{in}}^{r_{out}} \int_{0}^{\w} \frac{d \w '}{\dot{r}} dr .
\end{eqnarray}
Here we have used Hamilton's equation $\dot{r} = \frac{\partial H}{\partial p_{r}}$, and $H = M - \w$, where $\w$ is the energy of emitted particle, and $M$ mass of black hole.
By inserting Eq.(\ref{nullgeodesic}) into (\ref{action1}), and setting $r - r_H = \epsilon e^{i \theta}$, the above integral is performed on a semicircle centered at the real axis pole $r_H$,
\begin{eqnarray}
\label{action2}
\textrm{Im} (\mathcal{I}) & = & - \textrm{Im} \int_{r_{in}}^{r_{out}} \int_{0}^{\w} \frac{d \w ' d r}{(r - r_H) \; \k(M - \w')}  \nonumber \\
	  & = & - \pi \int_{0}^{\w} \frac{d \w '}{(r - r_H) \; \k (M - \w ')} .
\end{eqnarray}
It should be noticed that $r_{in} > r_{out}$ since horizon would shrink after emission.

According to corrections of surface gravity \cite{Zhang:2008rd,Banerjee:2008cf}, even in case of higher order quantum effects,
 Hawking temperature can still be expressed as $T = \frac{\k_{QG}}{2 \pi}$, where $\k_{QG}$ is quantum gravity surface gravity, with respect to classical $\k$.
Based on the first law of thermodynamics $d \w ' = d M ' = \frac{\k_{QG}}{2 \pi}$, we have
\begin{eqnarray*}
\textrm{Im} (\mathcal{I}) = - \frac{1}{2} \int_{S(M)}^{S(M - \w)} d S = \frac{1}{2} (S(M) - S(M - \w)) ,
\end{eqnarray*}
or more precisely
\begin{eqnarray}
 \label{deltaS}
\textrm{Im} (\mathcal{I}) = \frac{1}{2} (S_{QG} (M) - S_{QG} (M - \w)) ,
\end{eqnarray}
where $S_{QG}$ is the corrected area entropy for black hole
\begin{eqnarray}
 \label{SQG}
S_{QG} = \frac{A}{4 L_p ^2} + \alpha ln \frac{A}{L_p ^2} + \mathcal{O} (\frac{L_p ^2}{A} ) . 
\end{eqnarray}
This logarithmic correction is introduced both by string theory and loop quantum gravity \cite{Strominger:1996sh, Solodukhin:1997yy, Rovelli:1996dv, Ashtekar:1997yu,Kaul:2000kf},
in which the value of $\alpha$ is different. 
For Schwarzschild black hole, the area of horizon is $A = 4 \pi r_H ^2 = 16 \pi M^2$, and $L_p = \sqrt{\frac{\hbar G}{c^3}}$ is Planck length.
In units $G = c = k_B = \hbar = 1$, $L_p = 1$.
According to $\Gamma \sim exp(-2 \textrm{Im} \mathcal{I})$, the emission probability formula with quantum correction is
\begin{eqnarray}
 \label{probability}
\Gamma \sim exp( \Delta S_{QG}) = (1 - \frac{\w}{M})^{2 \alpha} \; exp \left(- 8 \pi \w (M - \frac{\w}{2}) \right) .
\end{eqnarray}
This expression is the basis of the following discussions.
Quantum gravity effects give an additional factor depending on the energy of emitted particle and the mass of black hole. 
The consequences of this factor is discussed in the following sections. 

It has been argued that the expression $\Gamma \sim exp(-2 \textrm{Im} \mathcal{I}) = exp (- 2 \textrm{Im} \int p_r d r )$ is not canonically invariant \cite{Akhmedov:2006un,Chowdhury:2006sk}.
However, it must be invariant in order to describe proper quantum mechanical observables.
This problem can be solved by using $\textrm{Im} \oint p_r d r$ instead of $2 \textrm{Im} \int p_r d r$. 
But using this canonically invariant formula leads to the so-called factor 2 problem, i.e., the temperature obtained is twice of the standard Hawking temperature \cite{Akhmedov:2006pg,Pilling:2007cn}.
The final solution is that, after adding the contribution of time variable transformation across the horizon \cite{Akhmedova:2008dz,Akhmedov:2008ru}, the correct temperature is recovered.
As pointed out in \cite{Pilling:2008xe}, the original expression $2 \textrm{Im} \int p_r d r$ gives the correct result in Painlev{\' e} coordinates,
 due to cancellation of the above two contributions. So we have used this original expression in our calculation.

\section{\label{sec3} Correlation between successive emissions} 

The expression (\ref{probability}), obtained by semi-classical tunneling method, shows deviation from thermal spectrum radiation. It is a result of back-reaction, 
after considering conservation of energy \cite{Parikh:2004ih}. Such a spectrum is an intriguing result, which may give some suggestions to the so-called black hole information loss paradox \cite{Hawking:1976ra}.
Many relevant discussions, e.g. \cite{Preskill:1992tc} , predicted on the idea of a purely thermal spectrum.

Consider two successive emissions, with energy $\w_1$ and $\w_2$ \cite{Parikh:2004rh}. The statistical correlation between quanta of hawking radiation is calculated, 
and the conclusion of trivial correlation is made in \cite{Arzano:2005rs}. However, we believe that, as in \cite{Zhang:2009jn}, the tunneling formulism, 
especially the deviation from thermal spectrum, should give some hints to the information paradox. Now we calculate the correlation using (\ref{probability}).

Firstly, if a black hole of initial mass $M$ emits a particle of energy $\w_1$, it follows that the associated probability is given by
\begin{eqnarray}
 \label{pbw1}
\Gamma(\w_1) = (1 - \frac{\w_1}{M})^{2 \alpha} \; exp \left(- 8 \pi \w_1 (M - \frac{\w_1}{2}) \right) .
\end{eqnarray}
The second emission, on the condition that the first one is $\w_1$, is
\begin{eqnarray}
 \label{pbw21}
\lefteqn{ \Gamma(\w_2 | \w_1)  =  (1 - \frac{\w_2}{M - \w_1})^{2 \alpha} } \nonumber \\
& & exp \left(- 8 \pi \w_2 (M - \w_1 - \frac{\w_2}{2}) \right) .
\end{eqnarray}
i.e., $\Gamma(\w_2 | \w_1)$ is the conditional probability \cite{Zhang:2009jn}. The emission of quanta $\w_1 + \w_2$
\begin{eqnarray}
 \label{pbw1w2}
\lefteqn{\Gamma(\w_1 + \w_2) = (1 - \frac{\w_1 + \w_2}{M})^{2 \alpha} } \nonumber \\
& & exp \left(- 8 \pi (\w_1 + \w_2) (M - \frac{\w_1 + \w_2}{2}) \right) .
\end{eqnarray}
We can check that $\Gamma(\w_1, \w_2) = \Gamma(\w_1) \Gamma(\w_2 | \w_1) = \Gamma(\w_1 + \w_2)$. 

The statistical correlation \cite{Parikh:2004rh} between emissions $\w_1$ and $\w_2$ is measured by
\begin{eqnarray*}
 \label{correlation}
\lefteqn{\chi (\w_1 + \w_2 ; \w_1, \w_2) } \nonumber \\
& & = ln \Gamma (\w_1 + \w_2) - ln \Gamma (\w_1) - ln \Gamma (\w_2) .
\end{eqnarray*}
In order to get $\Gamma (\w_2)$, we integrate the $\w_1$ variable in $\Gamma (\w_1, \w_2)$ and normalize it
\begin{eqnarray*}
\Lambda & = & \int_{0}^{M} \Gamma (\w) d \w , \\
 \Gamma (\w_2) & = & \frac{1}{\Lambda} \int_{0}^{M - \w_2} \Gamma (\w_1, \w_2) d \w_1  .
\end{eqnarray*}
Substituting (\ref{probability}) and (\ref{pbw1w2}) into the above expression, 
actually we need not do the integral, because the fraction is reducible. 
The result is 
\begin{eqnarray}
 \label{pbw2}
\Gamma(\w_2) = (1 - \frac{\w_2}{M})^{2 \alpha} \; exp \left(- 8 \pi \w_2 (M - \frac{\w_2}{2}) \right) .
\end{eqnarray}
 Now we can calculate the statistical correlation
\begin{eqnarray}
 \label{Newcorrelation}
\lefteqn{ \chi (\w_1 + \w_2 ; \w_1, \w_2) } \nonumber \\
& & = 8 \pi \w_1 \w_2 + 2 \alpha ln \left( 1 - \frac{\w_1 \w_2}{(M - \w_1)(M - \w_2)} \right).
\end{eqnarray}
This nontrivial result shows that subsequent emissions are statistically dependent, and correlations must exist between them. 
As discussed in \cite{Zhang:2009jn}, the amount of correlation hidden inside Hawking radiation is precisely equal to 
mutual information between the two sequential emissions. 
Our result is based on a tunneling formulism where energy conservation and back reaction are enforced. 
This nontrivial correlation plays an important role in considering the information paradox. 
It indicates that information would leak out during radiation.
In order to calculate entropy carried by Hawking radiation, the entire process of black hole evaporation should be considered.
After inclusion of the mutual information carried by radiation, conservation of entropy during black hole evaporation is checked in the next section.
However, even considering logarithmic corrections, tunneling formulism is still a semi-classical method, 
and the entire resolving of information paradox is dependent on a complete quantum gravity theory 
(e.g., details of how information is coded in the correlation should be explained).

Recently, nontrivial correlation is also obtained in \cite{Nozari:2008gp}, where Generalized Uncertainty Principle (GUP) and a modification of commutation relation is considered,
and similar conclusion of information leaking out is made there. 
However the influence of conditional probability is not considered in their calculation, 
and the nontrivial correlation entirely originates from the GUP effects.
So their discussion is not the same as ours.

\section{\label{sec4} Entropy conservation and black hole remnant}
 
Now let's consider black hole evaporation, based on the highly non-thermal spectrum (\ref{probability}).
The specific process is that particles $\w_1, \w_2, \dots, \w_n$ are successively emitted from the black hole. 
According to $\Gamma \sim e^{\Delta S}$, black hole gradually loses its entropy during evaporation.
The entropy is carried out both by the emitted energy and the correlations between them.
As suggested by \cite{Zhang:2009jn}, the total entropy carried out by radiation is
\begin{eqnarray}
\label{Scorrelation}
S(\w_1, \w_2, \dots, \w_n) & = & \sum_{i = 1}^n S(\w_i \; | \; \w_1, \w_2, \dots, \w_{i-1})  \nonumber \\
  & = & - ln \prod_{i = 1}^n \Gamma ( M - \sum_{j = 1}^{i - 1} \w_j \;| \; \w_i) ,
\end{eqnarray}
with
\begin{eqnarray*}
 \lefteqn{ \Gamma (\w_1)  =  (\frac{M - \w_1}{M})^{2 \alpha} \; exp \left(- 8 \pi \w_1 (M - \frac{\w_1}{2}) \right) , }  \\
\lefteqn{ \Gamma (\w_2 | \w_1) = (\frac{M - \w_1 - \w_2}{M - \w_1})^{2 \alpha} } \\
& & exp \left(- 8 \pi \w_2 (M - \w_1 - \frac{\w_1}{2}) \right) , \\
 & \dots &  , \\
 \lefteqn{ \Gamma  (\w_n | \w_1, \w_2 , \dots, \w_{n-1}) = (\frac{\w_c}{\w_n + \w_c})^{2 \alpha} } \\
& & exp \left(- 8 \pi \w_n (\w_n + \w_c - \frac{\w_n}{2}) \right) ,
\end{eqnarray*}
and $\w_1 + \w_2 + \dots + \w_n + \w_c = M$, $\w_c$ is black hole remnant. Then, 
\begin{eqnarray}
 \label{rmentropy}
S & = & - ln \left( (\frac{\w_c}{M})^{2 \alpha} \; exp ( - 4 \pi (M^2 - \w_c ^2)) \right) \nonumber \\
  & = & 4 \pi M^2 + \alpha ln \frac{M^2}{\w_c ^2} - 4 \pi \w_c ^2 .
\end{eqnarray}
We must emphasize that the existence of $\w_c$ is essential in the above calculation. 
Otherwise, if $\w_c = 0$, the whole $\Gamma$-products in (\ref{Scorrelation}) would be zero, and the entropy would be divergent!
The origin of this divergent is that the quantum gravity corrected entropy (\ref{SQG}) is not valid for an infinitesimal black hole. 
In string theory the sign of $\alpha$ depends on the number of field species appearing in the low energy approximation \cite{Solodukhin:1997yy}. 
In the case of loop quantum gravity $\alpha$ is a negative coefficient, whose value has been rigorously fixed at $\alpha = - \frac{1}{2}$ \cite{Meissner:2004ju}. 
Assume that $\alpha < 0$, a black hole remnant $\w_c = \sqrt{\frac{-\alpha}{4 \pi}}$ is suggested and argued in \cite{Li:2007ga}.
This remnant can also be obtained by demanding the critical mass given by (\ref{SQG}), on the condition that black hole entropy is monotonic increasing with its mass, 
i.e., $\frac{\partial{S_{QG}}}{\partial{M}}|_{M = \w_c} = 0$. 
In classical cases black hole would evaporate completely, and the entropy is carried out entirely by radiation.
After including quantum gravity effects, the appearance of black hole remnant is natural,
 since generalized uncertainty principle may prevent black hole evaporating completely\cite{Adler:2001vs}.
The idea of a black hole remnant also comes from non-commutative geometry which introduces a minimal length via a non-trivial commutation relation between coordinates \cite{Nicolini:2005vd,Ansoldi:2006vg,Nicolini:2008aj}.
But in the absence of a well-defined quantum gravity theory, an exact formula of remnant is unavailable.

The Bekenstein-Hawking Entropy of black hole exactly saturates the Bekenstein's entropy bound \cite{Bekenstein:1980jp}.
According to \cite{Bekenstein:1993dz}, Bekenstein's entropy bound should also be applied to the remnant. 
We assume the remnant is something that has black hole properties and also saturates the bound. 
Therefore the entropy of remnant should have a similar form as a black hole. With logarithmic correction,  $S_c = 4 \pi \w_c ^2 + \alpha ln 16 \pi \w_c ^2$.
Eq.(\ref{rmentropy}) can further be expressed as
\begin{eqnarray}
 \label{entropyconservation}
S & = & 4 \pi M^2 + \alpha ln \frac{16 \pi M^2}{16 \pi \w_c ^2} - 4 \pi \w_c ^2 \nonumber \\
  & = & (\frac{A}{4} + \alpha ln A) - (4 \pi \w_c ^2 + \alpha ln 16 \pi \w_c ^2) \nonumber \\
  & = & S_{QG} - S_c . 
\end{eqnarray}
We interpret this formula as conservation of entropy, which means that total entropy of original black hole is equal to the addition of entropy carried out by radiation 
and entropy of black hole remnant. 
In a recent paper, the idea of entropy conservation is also found on tunneling formulism applied to FRW cosmology model \cite{Zhu:2009wa}.
Our result, together with \cite{Zhang:2009jn}, implies that in considering Hawking radiation as a tunneling process, 
no information loss occurs, and therefore black hole evaporation is a unitary process.

\section{Summary}\

Using tunneling formulism and quantum gravity corrected entropy, the modified radiation probability is derived.
Based on this probability, we have discussed the correlation between successively emitted particles. 
Black hole evaporation process is considered and conservation of entropy is checked.
The role of black hole remnant is important in considering this process, otherwise the entropy would be divergent.
We conclude that, in the case of quantum gravity corrections, the information loss paradox can also be explained, 
and unitarity of black hole evaporation process can be preserved. 



\section*{Acknowledgement}

We would like to thank C. Cao, Q. J. Cao, Y. J. Du, J. L. Li, Q. Ma, Y. Q. Wang and Y. Xiao
for useful discussions. The work is supported in part by the NNSF of
China Grant No. 90503009, No. 10775116, and 973 Program Grant No.
2005CB724508.


\begin{thebibliography}{00}


%\cite{Hawking:1974sw}
\bibitem{Hawking:1974sw}
  S.~W.~Hawking,
  %``Particle Creation By Black Holes,''
  Commun.\ Math.\ Phys.\  {\bf 43}, 199 (1975)
  [Erratum-ibid.\  {\bf 46}, 206 (1976)].
  %%CITATION = CMPHA,43,199;%%

%
\bibitem{thermodynamics} J. M. Bardeen, B. Carter and S. W. Hawking, 
Commun. Math. Phys. {\bf 31}, 161 (1973); J. D. Beckenstein, 
Phys. Rev. D {\bf 7}, 2233 (1973); 
Robert M. Wald, Phys. Rev. D {\bf 20}, 1271 (1979).

%\cite{Hawking:1976ra}
\bibitem{Hawking:1976ra}
  S.~W.~Hawking,
  %``Breakdown Of Predictability In Gravitational Collapse,''
  Phys.\ Rev.\  D {\bf 14}, 2460 (1976).
  %%CITATION = PHRVA,D14,2460;%%

%\cite{Russo:2005aw}
\bibitem{Russo:2005aw}
  J.~G.~Russo,
  %``The information problem in black hole evaporation: Old and recent
  %results,''
  arXiv:hep-th/0501132.
  %%CITATION = HEP-TH/0501132;%%

%\cite{Srinivasan:1998ty}
\bibitem{Srinivasan:1998ty}
  K.~Srinivasan and T.~Padmanabhan,
  %``Particle production and complex path analysis,''
  Phys.\ Rev.\  D {\bf 60}, 024007 (1999)
  [arXiv:gr-qc/9812028].
  %%CITATION = PHRVA,D60,024007;%%

%\cite{Parikh:1999mf}
\bibitem{Parikh:1999mf}
  M.~K.~Parikh and F.~Wilczek,
  %``Hawking radiation as tunneling,''
  Phys.\ Rev.\ Lett.\  {\bf 85}, 5042 (2000)
  [arXiv:hep-th/9907001].
  %%CITATION = PRLTA,85,5042;%%

%\cite{Shankaranarayanan:2000qv}
\bibitem{Shankaranarayanan:2000qv}
  S.~Shankaranarayanan, T.~Padmanabhan and K.~Srinivasan,
  %``Hawking radiation in different coordinate settings: Complex paths
  %approach,''
  Class.\ Quant.\ Grav.\  {\bf 19}, 2671 (2002)
  [arXiv:gr-qc/0010042].
  %%CITATION = CQGRD,19,2671;%%


%\cite{Zhang:2009jn}
\bibitem{Zhang:2009jn}
  B.~Zhang, Q.~y.~Cai, L.~You and M.~S.~Zhan,
  %``Hidden Messenger Revealed in Hawking Radiation: a Resolution to the Paradox
  %of Black Hole Information Loss,''
  Phys.\ Lett.\  B {\bf 675}, 98 (2009)
  arXiv:0903.0893 [hep-th].
  %%CITATION = ARXIV:0903.0893;%%

%\cite{Arzano:2005fu}
\bibitem{Arzano:2005fu}
  M.~Arzano,
  %``Information leak through the quantum horizon,''
  Mod.\ Phys.\ Lett.\  A {\bf 21}, 41 (2006)
  [arXiv:hep-th/0504188].
  %%CITATION = MPLAE,A21,41;%%

%\cite{Arzano:2005rs}
\bibitem{Arzano:2005rs}
  M.~Arzano, A.~J.~M.~Medved and E.~C.~Vagenas,
  %``Hawking radiation as tunneling through the quantum horizon,''
  JHEP {\bf 0509}, 037 (2005)
  [arXiv:hep-th/0505266].
  %%CITATION = JHEPA,0509,037;%%

\bibitem{temperatures} 
E.C. Vagenas, Phys. Lett. B {\bf 533}, 302 (2002); \\
A.J.M. Medved and E.C. Vagenas, Mod. Phys. Lett. A {\bf 20}, 2449 (2005); \\
Michele Arzano, A.J.M. Medved and E.C. Vagenas, JHEP 0509, 037 (2005); \\
Qing-Quan Jiang, Shuang-Qing Wu, and Xu Cai, 
Phys. Rev. D {\bf 73}, 064003 (2006); \\
Yapeng Hu, Jingyi Zhang, and Zheng Zhao, gr-qc/0611026 and gr-qc/0611085; \\
Zhibo Xu and Bin Chen, Phys. Rev. D {\bf 75}, 024041 (2007); \\
Xiaoning Wu and Sijie Gao, Phys. Rev. D {\bf 75}, 044027 (2007); \\
Cheng-Zhou Liu and Jian-Yang Zhu, gr-qc/0703055; \\
Ryan Kerner and R. B. Mann, Phys. Rev. D {\bf 75}, 084022 (2007); \\
Bhramar Chatterjee, A. Ghosh, and P. Mitra, hep-th/0704.1746; \\
Ji-Rong Ren, Ran Li, and Fei-Hu Liu, Mod.\ Phys.\ Lett.\  A {\bf 23}, 3419 (2009), gr-qc/0705.4336
%

%\cite{Zhang:2008rd}
\bibitem{Zhang:2008rd}
  B.~Zhang, Q.~y.~Cai and M.~s.~Zhan,
  %``Hawking radiation as tunneling derived from Black Hole Thermodynamics
  %through the quantum horizon,''
  Phys.\ Lett.\  B {\bf 665}, 260 (2008)
  [arXiv:0806.2015 [hep-th]].
  %%CITATION = PHLTA,B665,260;%%

%\cite{Pilling:2007cn}
\bibitem{Pilling:2007cn}
  T.~Pilling,
  %``Tunneling derived from Black Hole Thermodynamics,''
  Phys.\ Lett.\  B {\bf 660}, 402 (2008)
  [arXiv:0709.1624 [gr-qc]].
  %%CITATION = PHLTA,B660,402;%%

%\cite{Banerjee:2008cf}
\bibitem{Banerjee:2008cf}
  R.~Banerjee and B.~R.~Majhi,
  %``Quantum Tunneling Beyond Semiclassical Approximation,''
  JHEP {\bf 0806}, 095 (2008)
  [arXiv:0805.2220 [hep-th]].
  %%CITATION = JHEPA,0806,095;%%

%\cite{Strominger:1996sh}
\bibitem{Strominger:1996sh}
A.~Strominger and C.~Vafa,
%``Microscopic Origin of the Bekenstein-Hawking Entropy,''
Phys.\ Lett.\ B {\bf 379}, 99 (1996)
[arXiv:hep-th/9601029];

%\cite{Solodukhin:1997yy}
\bibitem{Solodukhin:1997yy}
S.~N.~Solodukhin,
%``Entropy of Schwarzschild black hole and string-black hole  correspondence,''
Phys.\ Rev.\ D {\bf 57}, 2410 (1998)
[arXiv:hep-th/9701106].


%\cite{Rovelli:1996dv}
\bibitem{Rovelli:1996dv}
C.~Rovelli,
%``Black hole entropy from loop quantum gravity,''
Phys.\ Rev.\ Lett.\  {\bf 77}, 3288 (1996)
[arXiv:gr-qc/9603063].


%\cite{Ashtekar:1997yu}
\bibitem{Ashtekar:1997yu}
A.~Ashtekar, J.~Baez, A.~Corichi and K.~Krasnov,
%``Quantum geometry and black hole entropy,''
Phys.\ Rev.\ Lett.\  {\bf 80}, 904 (1998)
[arXiv:gr-qc/9710007].

%\cite{Kaul:2000kf}
\bibitem{Kaul:2000kf}
R.~K.~Kaul and P.~Majumdar,
%``Logarithmic correction to the Bekenstein-Hawking entropy,''
Phys.\ Rev.\ Lett.\  {\bf 84}, 5255 (2000)
[arXiv:gr-qc/0002040].

%\cite{Akhmedov:2006un}
\bibitem{Akhmedov:2006un}
  E.~T.~Akhmedov, V.~Akhmedova, T.~Pilling and D.~Singleton,
  %``Thermal radiation of various gravitational backgrounds,''
  Int.\ J.\ Mod.\ Phys.\  A {\bf 22}, 1705 (2007)
  [arXiv:hep-th/0605137].
  %%CITATION = IMPAE,A22,1705;%%

%\cite{Chowdhury:2006sk}
\bibitem{Chowdhury:2006sk}
  B.~D.~Chowdhury,
  %``Problems with Tunneling of Thin Shells from Black Holes,''
  Pramana {\bf 70}, 593 (2008)
  [Pramana {\bf 70}, 3 (2008)]
  [arXiv:hep-th/0605197].
  %%CITATION = PRAMC,70,3;%%

%\cite{Akhmedov:2006pg}
\bibitem{Akhmedov:2006pg}
  E.~T.~Akhmedov, V.~Akhmedova and D.~Singleton,
  %``Hawking temperature in the tunneling picture,''
  Phys.\ Lett.\  B {\bf 642}, 124 (2006)
  [arXiv:hep-th/0608098].
  %%CITATION = PHLTA,B642,124;%%

%\cite{Akhmedova:2008dz}
\bibitem{Akhmedova:2008dz}
  V.~Akhmedova, T.~Pilling, A.~de Gill and D.~Singleton,
  %``Temporal contribution to gravitational WKB-like calculations,''
  Phys.\ Lett.\  B {\bf 666}, 269 (2008)
  [arXiv:0804.2289 [hep-th]].
  %%CITATION = PHLTA,B666,269;%%

%\cite{Akhmedov:2008ru}
\bibitem{Akhmedov:2008ru}
  E.~T.~Akhmedov, T.~Pilling and D.~Singleton,
  %``Subtleties in the quasi-classical calculation of Hawking radiation,''
  Int.\ J.\ Mod.\ Phys.\  D {\bf 17}, 2453 (2008)
  [arXiv:0805.2653 [gr-qc]].
  %%CITATION = IMPAE,D17,2453;%%

%\cite{Pilling:2008xe}
\bibitem{Pilling:2008xe}
  T.~Pilling,
  %``Quasi-classical Hawking Temperatures and Black Hole Thermodynamics,''
  arXiv:0809.2701 [hep-th].
  %%CITATION = ARXIV:0809.2701;%%

%\cite{Parikh:2004ih}
\bibitem{Parikh:2004ih}
  M.~K.~Parikh,
  %``A secret tunnel through the horizon,''
  Int.\ J.\ Mod.\ Phys.\  D {\bf 13}, 2351 (2004)
  [Gen.\ Rel.\ Grav.\  {\bf 36}, 2419 (2004)]
  [arXiv:hep-th/0405160].
  %%CITATION = GRGVA,36,2419;%%

%\cite{Preskill:1992tc}
\bibitem{Preskill:1992tc}
  J.~Preskill,
  %``Do black holes destroy information?,''
  arXiv:hep-th/9209058.
  %%CITATION = HEP-TH/9209058;%%

%\cite{Parikh:2004rh}
\bibitem{Parikh:2004rh}
  M.~K.~Parikh,
  %``Energy conservation and Hawking radiation,''
  arXiv:hep-th/0402166.
  %%CITATION = HEP-TH/0402166;%%

%\cite{Nozari:2008gp}
\bibitem{Nozari:2008gp}
  K.~Nozari and S.~Hamid Mehdipour,
  %``Quantum Gravity and Recovery of Information in Black Hole Evaporation,''
  Europhys.\ Lett.\  {\bf 84}, 20008 (2008)
  [arXiv:0804.4221 [gr-qc]].
  %%CITATION = EULEE,84,20008;%%

%\cite{Meissner:2004ju}
\bibitem{Meissner:2004ju}
  K.~A.~Meissner,
  %``Black hole entropy in loop quantum gravity,''
  Class.\ Quant.\ Grav.\  {\bf 21}, 5245 (2004)
  [arXiv:gr-qc/0407052].
  %%CITATION = CQGRD,21,5245;%%

%\cite{Li:2007ga}
\bibitem{Li:2007ga}
  X.~Li,
  %``A note on the black hole remnant,''
  Phys.\ Lett.\  B {\bf 647} (2007) 207.
  %%CITATION = PHLTA,B647,207;%%

%\cite{Adler:2001vs}
\bibitem{Adler:2001vs}
  R.~J.~Adler, P.~Chen and D.~I.~Santiago,
  %``The generalized uncertainty principle and black hole remnants,''
  Gen.\ Rel.\ Grav.\  {\bf 33}, 2101 (2001)
  [arXiv:gr-qc/0106080].
  %%CITATION = GRGVA,33,2101;%%

%\cite{Nicolini:2005vd}
\bibitem{Nicolini:2005vd}
  P.~Nicolini, A.~Smailagic and E.~Spallucci,
  %``Noncommutative geometry inspired Schwarzschild black hole,''
  Phys.\ Lett.\  B {\bf 632}, 547 (2006)
  [arXiv:gr-qc/0510112].
  %%CITATION = PHLTA,B632,547;%%

%\cite{Ansoldi:2006vg}
\bibitem{Ansoldi:2006vg}
  S.~Ansoldi, P.~Nicolini, A.~Smailagic and E.~Spallucci,
  %``Noncommutative geometry inspired charged black holes,''
  Phys.\ Lett.\  B {\bf 645}, 261 (2007)
  [arXiv:gr-qc/0612035].
  %%CITATION = PHLTA,B645,261;%%

%\cite{Nicolini:2008aj}
\bibitem{Nicolini:2008aj}
  P.~Nicolini,
  %``Noncommutative Black Holes, The Final Appeal To Quantum Gravity: A
  %Review,''
  Int.\ J.\ Mod.\ Phys.\  A {\bf 24}, 1229 (2009)
  [arXiv:0807.1939 [hep-th]].
  %%CITATION = IMPAE,A24,1229;%%

%\cite{Bekenstein:1980jp}
\bibitem{Bekenstein:1980jp}
  J.~D.~Bekenstein,
  %``A Universal Upper Bound On The Entropy To Energy Ratio For Bounded
  %Systems,''
  Phys.\ Rev.\  D {\bf 23}, 287 (1981).
  %%CITATION = PHRVA,D23,287;%%

%\cite{Bekenstein:1993dz}
\bibitem{Bekenstein:1993dz}
  J.~D.~Bekenstein,
  %``Entropy Bounds And Black Hole Remnants,''
  Phys.\ Rev.\  D {\bf 49}, 1912 (1994)
  [arXiv:gr-qc/9307035].
  %%CITATION = PHRVA,D49,1912;%%

%\cite{Zhu:2009wa}
\bibitem{Zhu:2009wa}
  T.~Zhu, J.~R.~Ren and D.~Singleton,
  %``Hawking-like radiation as tunneling from the apparent horizon in a FRW
  %Universe,''
  arXiv:0902.2542 [hep-th].
  %%CITATION = ARXIV:0902.2542;%%


\end{thebibliography}
\end{document}